\documentclass[a4paper,11pt]{article}
\usepackage{pos}

\rightline{
	{\vbox {			
		\hbox{JLAB-THY-24-4144}
}}}
\vspace*{-0.6cm}
\title{Factorized QED Contribution to Lepton-Hadron DIS}
\author[a]{Justin Cammarota}
\author*[a,b]{Jian-Wei Qiu}
\author[c]{Kazuhiro Watanabe}
\author[b]{Jia-Yue Zhang}

\affiliation[a]{Department of Physics, William \& Mary, 
Williamsburg, Virginia 23187, USA}

\affiliation[b]{Theory Center, Jefferson Lab,
Newport News, Virginia 23606, USA}

\affiliation[c]{Faculty of Science and Technology, 
Seikei University, Musashino, 
Tokyo 180-8633, Japan}

\emailAdd{jcammarota@wm.edu}
\emailAdd{jqiu@jlab.org}
\emailAdd{kazuhr.watanabe@gmail.com}
\emailAdd{jzhang@jlab.org}

\abstract{We present the first calculation of next-to-leading order (NLO) factorized QED contributions to the short-distance hard coefficients of inclusive lepton-hadron deep inelastic scattering (DIS) in a joint QCD and QED factorization approach.  We demonstrate how the joint factorization consistently factorize all perturbative collinear sensitivities of partonic scattering in both QCD and QED into corresponding universal hadron and lepton distribution functions without the need of any parameters other than the standard factorization scale.  We discuss the necessary modification to DGLAP-type evolution of the parton and lepton distribution functions in this joint factorization approach.  We also discuss the potential impact of this joint factorization approach on the extraction of partonic information from lepton-hadron DIS.}

\FullConference{31st International Workshop on Deep Inelastic Scattering (DIS2024)\\
 8–12 April 2024\\
Grenoble, France\\}

\begin{document}
\maketitle
%==================================================================

%==================================================================
\section{Introduction}
\label{sec1:intro}

The lepton-hadron deep-inelastic scattering (DIS) has played an important role in the discovery and development of the fundamental theory of strong interactions known as Quantum Chromodynamics (QCD) (for reviews, see~\cite{Brambilla:2014jmp,Gross:2022hyw}). It can also provide unique opportunities to help probe and explore the internal structure of the colliding hadron, as well as the emergence of hadrons from produced partons in the scattering~\cite{Accardi:2012qut}.  With a colliding lepton of momentum $\ell$ and a hadron of momentum $P$, measuring the scattered lepton of momentum $\ell'$ can provide us a hard probe from the large momentum transfer $Q^2=-(\ell-\ell')^2 \gg \Lambda_{\rm QCD}^2$, allowing us to apply QCD factorization~\cite{Collins:1989gx} to connect the measurements of hadrons to information about the quarks and gluons inside them.  

With a large momentum transfer $Q^2$, the inclusive scattering necessarily generates both QCD and QED collision-induced radiation, affecting the measured cross sections.  The QCD factorization provides a systematic way to evaluate the contributions from the collision-induced QCD radiation.  On the other hand, historically, QED contributions have been evaluated perturbatively by calculating collision-induced radiation from observed leptons as radiative corrections (RCs) to the lowest order Born cross section in QED~\cite{Mo:1968cg,Bardin:1989vz,Badelek:1994uq,Kripfganz:1990vm,Spiesberger:1994dm,Blumlein:2002fy}.  
Such an approach often needs to introduce unknown parameter(s) that could affect the predictive power of RCs.  When we look into the final-state of the lepton-hadron scattering, the approach of RCs could limit our capability to use the technique of azimuthal modulations to help separate information on different transverse momentum dependent PDFs (or simply, TMDs) from semi-inclusive DIS (or SIDIS)~\cite{Liu:2020rvc}.  A new joint QCD and QED factorization approach was recently proposed to treat the collision-induced QCD and QED radiation equally~\cite{Liu:2021jfp}. 

In this talk, we present the first calculation of fully factorized next-to-leading order (NLO) QED contributions to the short-distance hard coefficients of inclusive lepton-hadron DIS in the joint QCD and QED factorization approach.  Unlike the calculation of RCs in the literature, our results are completely infrared safe, only depending on the standard factorization scale, while all leading power non-perturbative and collinear sensitive radiative contributions are included in universal lepton distribution functions (LDFs) and/or lepton fragmentation functions (LFFs).  Because of the nonperturbative nature of collision-induced radiation, we explore the uncertainties of the traditional perturbative approach for RCs and corresponding impact on the precision of DIS.

%==================================================================
\section{Factorization Formalism and QED Contributions}
\label{sec2:qed}

The inclusive lepton-hadron DIS, $e(\ell)+h(P)\to e'(\ell') +X$, is effectively an inclusive cross section for producing a lepton of momentum $\ell'$.  At the lowest order (Born cross section) in QED coupling, the scattering is approximated by one-photon exchange between the colliding lepton and hadron.  The corresponding momentum transfer by the exchanged photon, $q=\ell - \ell'$ with $Q^2=-q^2$, serves as a hard probe for exploring partonic structure of colliding hadron.  However, collision-induced photon radiation can change the momentum transfer between the colliding lepton and hadron so as the precision of the hard probe~\cite{Liu:2021jfp}.  

Instead of relying on perturbative calculation of QED radiative corrections, which are collinear sensitive and potentially nonperturbative since a radiated photon from the colliding lepton can turn into a light quark-antiquark pair at a nonperturbative scale, we evaluate the inclusive DIS cross section in a joint QCD and QED factorization approach to absorb all collinear sensitive and process-independent nonperturbative contributions into LDFs, LFFs and PDFs, respectively, in terms of the following factorization formalism~\cite{Liu:2021jfp} 
\begin{eqnarray}
E'\frac{d\sigma_{\ell P\to \ell' X}}{d^3\ell'} 
& \approx  & 
\frac{1}{2s} \sum_{ija} \int_{\zeta_{\rm min}}^1 \frac{d\zeta}{\zeta^2} \, D_{e/j}(\zeta,\mu^2) 
 \int_{\xi_{\rm min}}^1 \frac{d\xi}{\xi} \, f_{i/e}(\xi,\mu^2) 
 \label{eq:fac} \\
 \nonumber
 &  & \times
 \int_{x_{\rm min}}^1 \frac{dx}{x} \, f_{a/h}(x,\mu^2) \,
 \widehat{H}_{ia\to j X}(\xi\ell, xP,\ell'/\zeta,\mu^2) 
 + {\cal O}\left(1/\ell'_{T}\right)^\alpha
\end{eqnarray}
where $f_{i/e}$, $D_{e/j}$ and $f_{a/h}$ are the LDFs, LFFs and PDFs, respectively, $\mu$ represents the factorization scale (chosen to be the same as the renormalization scale), and short-distance hard parts
$\widehat{H}_{ia\to j X}$ are perturbatively calculable in a power series of $(\alpha_{em}^{m}\alpha_s^n)$ with $m\geq 2$ and $n\geq 0$.  
The lowest order hard part is given by~\cite{Liu:2021jfp}
\begin{equation}
 \widehat{H}^{(2,0)}_{eq\to eX}
 = e_q^2 \left(4\alpha_{em}^2\right) \frac{x^2\zeta\left[(\zeta\xi s)^2+u^2\right]}{(\xi t)^2 \, (\zeta\xi s+u)}\,  \delta(x-x_{\rm min})
 \label{eq:H20}
 \end{equation}
 where $s=(\ell+P)^2$, $u=(\ell'-P)^2=(y-1)s$, $t=(\ell-\ell')^2=-Q^2$, and 
 \begin{equation}
 x_{\rm min} = \frac{\xi \, x_B \, y}{\xi\zeta +y -1}, \quad
 \xi_{\rm min} = \frac{1-y}{\zeta-x_B y}, \quad
 \zeta_{\rm min} = 1- (1-x_B)y, \quad \mbox{and} \quad
 x_B=\frac{Q^2}{2P\cdot q}\, .
 \label{eq:limits}
 \end{equation}
In Eq.~(\ref{eq:fac}), the factorization formula depends on universal PDFs of hadron $h$, not the DIS structure functions of $h$, and is valid beyond the traditional one-photon approximation for DIS. 

%----------------------------------------------------------------
% Figure: NLO Feynman diagrams
%----------------------------------------------------------------
\begin{figure}[htbp]
	\centering
	\begin{tabular}{cc}
	     \begin{minipage}[t]{0.64\textwidth}
	          \centering
		 \includegraphics[width=0.8\textwidth]{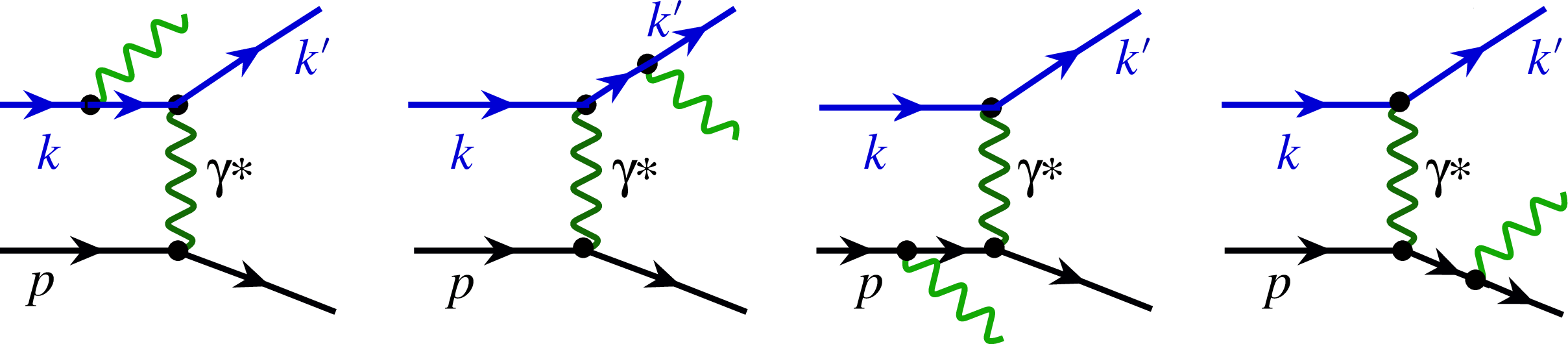} \\ \vskip 0.1in
		 \includegraphics[width=0.9\textwidth]{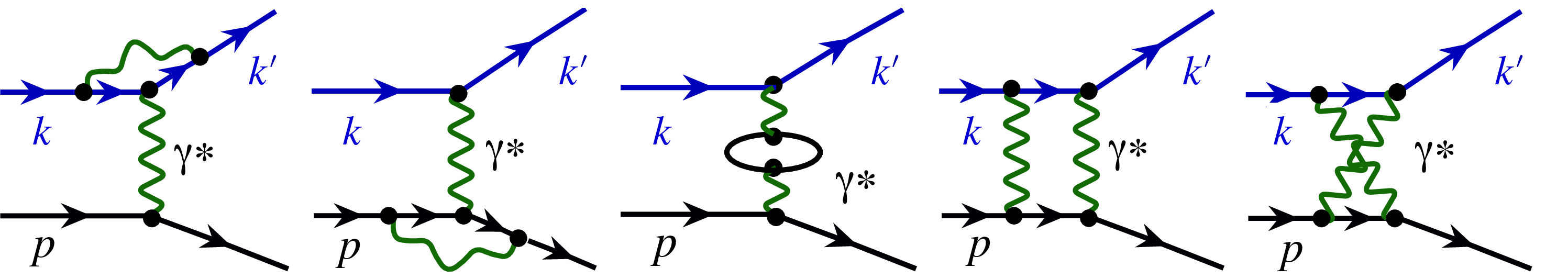}
	     \end{minipage}
		&{\hskip 0.05in}
	     \begin{minipage}[t]{0.28\textwidth}
	          \centering
		  \includegraphics[width=0.52\textwidth]{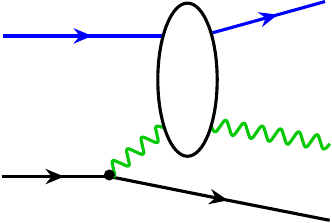} \\ \vskip 0.18in
		  \includegraphics[width=0.48\textwidth]{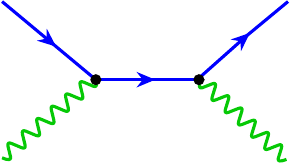} 
		        {\hskip 0.04\textwidth} 
		  \includegraphics[width=0.42\textwidth]{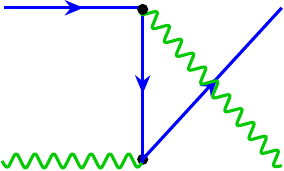} 
	     \end{minipage} \\
		(a) &{\hskip 0.05in} (b)
	\end{tabular}
	\caption{(a) NLO Feynman diagrams for real (top) and virtual (bottom) QED contributions to $\sigma_{eq\to eX}^{(3,0)}$ in Eq.~(\ref{eq:H30}).
	(b) Diagram associated with collinear photon radiation from colliding quark line (top) and lowest order diagrams for the hard part (bottom).
	              }
	\label{fig:feyn_nlo}
\end{figure}
%----------------------------------------------------------------

To compute the next-to-leading order infrared-safe hard part in Eq.~(\ref{eq:fac}) in QED, we apply the factorization formalism to a quark state and expand it to the order $\alpha_{em}^3$~\cite{Liu:2021jfp} and obtain, 
\begin{equation}
\widehat{H}^{(3,0)}_{eq\to eX} =
\sigma_{eq\to eX}^{(3,0)} - 
D^{(1)}_{e/e}\otimes_{\zeta} \widehat{H}^{(2,0)}_{eq\to eX} - 
f^{(1)}_{e/e}\otimes_{\xi} \widehat{H}^{(2,0)}_{eq\to eX} -
f^{(1)}_{q/q}\otimes_{x} \widehat{H}^{(2,0)}_{eq\to eX} -
f^{(1)}_{\gamma/q}\otimes_{x} \widehat{H}^{(2,0)}_{e\gamma\to eX} 
\label{eq:H30}
\end{equation}
where $\sigma_{eq\to eX}^{(3,0)}=2sE'd\sigma^{(3,0)}_{eq\to eX}/d^3\ell'$ represents the partonic cross section at ${\cal O}(\alpha_{em}^{3})$, whose real contribution is given by the diagrams in the top row of Fig.~\ref{fig:feyn_nlo}(a) and virtual contribution is given by the one-loop diagrams in the bottom row of Fig.~\ref{fig:feyn_nlo}(a), the $\otimes_i$ represents the convolution of variable $i=\zeta,\xi,x$ in Eq.~(\ref{eq:fac}) while setting the other variables ($\neq i$) to be 1 in $\widehat{H}^{(2,0)}$.  The $\sigma_{eq\to eX}^{(3,0)}$ in Eq.~(\ref{eq:H30}) is derived by calculating the diagrams in Fig.~\ref{fig:feyn_nlo}(a) with dimensional regularization and $\overline{\rm MS}$ factorization scheme.  It has collinear divergences caused by real photon radiation from observed and charged fermions - the observed electron plus the colliding electron and quark.  These divergences are exactly canceled by the first three subtraction terms, respectively in Eq.~(\ref{eq:H30}).  The $\sigma_{eq\to eX}^{(3,0)}$ has an additional collinear divergence when the momentum of the exchanged virtual photon goes on-shell and parallel to colliding quark while the momentum $\ell'$ of the large angle scattered lepton is balanced by the momentum of the radiated real photon, which can be represented by the top diagram in Fig.~\ref{fig:feyn_nlo}(b).  This collinear divergence of $\sigma_{eq\to eX}^{(3,0)}$ is often not included in the calculation of RCs when the exchanged photon is kept far off-shell.  In our joint QCD and QED factorization approach, this collinear divergence of $\sigma_{eq\to eX}^{(3,0)}$ is exactly canceled by the last subtraction term in Eq.~(\ref{eq:H30}) with the leading order hard part $\widehat{H}^{(2,0)}_{e\gamma\to eX}$ given by the $2\to 2$ real diagrams in the bottom row of Fig.~\ref{fig:feyn_nlo}(b).  The complete, analytical and finite result for $\widehat{H}^{(3,0)}_{eq\to eX}$ is presented in Ref.~\cite{CQWZ:nlo:qed}.  

The derivation of the $\widehat{H}^{(3,0)}_{eq\to eX}$ confirms that the flavor indices $i,j,a$ in Eq.~(\ref{eq:fac}) should include both QCD and QED ``parton-flavors". The evolution equations for LDFs, LFFs and PDFs will have kernels to mix QCD and QED partons since quarks and antiquarks carry $em$ charges.  Consequently, the LDFs and LFFs are effectively nonperturbative, since photon can be transformed into a light quark-antiquark pair at a nonperturbative scale, but, universal and can be systematically extracted from data through a joint QCD and QED global analysis~\cite{QW:evo}. 

%==================================================================
\section{Numerical Impact}
\label{sec3:numerical}

We study the impact of the collision-induced QED radiation to the inclusive lepton-hadron DIS by evaluating the factorized DIS cross sections with various LDFs, LFFs, and the LO and NLO hard parts in QED.  We used CTEQ CT18 unpolarized PDFs~\cite{Hou:2019efy} and did not see much difference from using other set of PDFs~\cite{CQWZ:nlo:qed}.  

We define the DIS cross section {\it without} the collision-induced QED radiation, $d\sigma_{\ell P\to \ell' X}^{\rm LO-NR}$, by setting $D_{e/j}(\zeta)=\delta(1-\zeta)\delta^{ej}$ and $f_{i/e}=\delta(1-\xi)\delta^{ie}$, and 
$\widehat{H}_{ia\to j X} =\widehat{H}^{(2,0)}_{ea\to eX}$ with $a$ of $\sum_a$ covers only QCD partons in Eq.~(\ref{eq:fac}).  We define the DIS cross section with perturbative QED RCs, $d\sigma_{\ell P\to \ell' X}^{\rm LO-Pert}$, by using LDFs and LFFs calculated in perturbative QED to NLO~\cite{Liu:2021jfp}
\begin{eqnarray}
f^{(\rm NLO)}_{e/e}(\xi,\mu^2) &=& 
 \delta(1-\xi) + \frac{ \alpha_{em}}{2\pi} \left[\frac{1+\xi^2}{1-\xi}\ln\frac{\mu^2}{(1-\zeta)^2m_e^2}\right]_+ \, ,
\label{eq:ldf} \\
D^{(\rm NLO)}_{e/e}(\zeta,\mu^2) &=& 
 \delta(1-\zeta) + \frac{ \alpha_{em}}{2\pi} \left[\frac{1+\zeta^2}{1-\zeta}\ln\frac{\zeta^2\mu^2}{(1-\zeta)^2m_e^2}\right]_+ \, ,
\label{eq:lff} 
\end{eqnarray}
with the electron mass as a cut off for the perturbative collinear singularity in QED.  

\begin{figure}[htbp]
	\centering
	\begin{tabular}{cc}
		\includegraphics[scale=0.46]{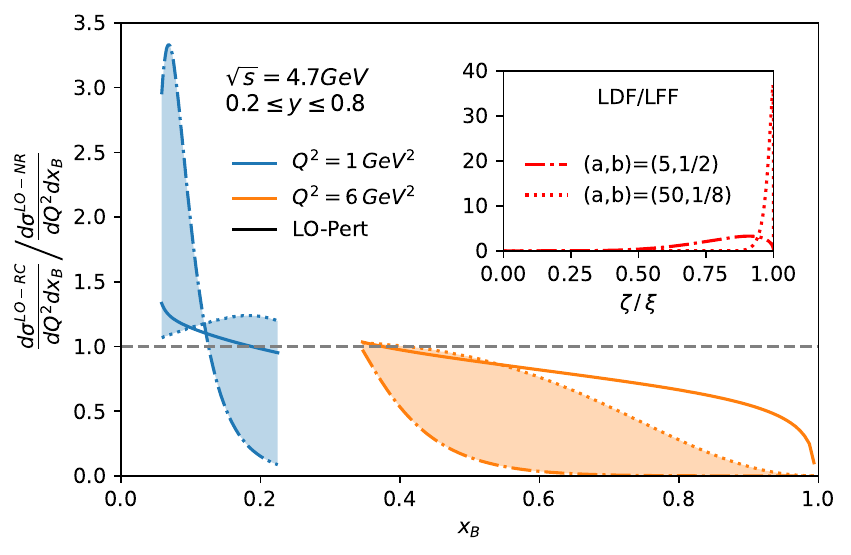} 
		&{\hskip 0.1in}
		\includegraphics[scale=0.46]{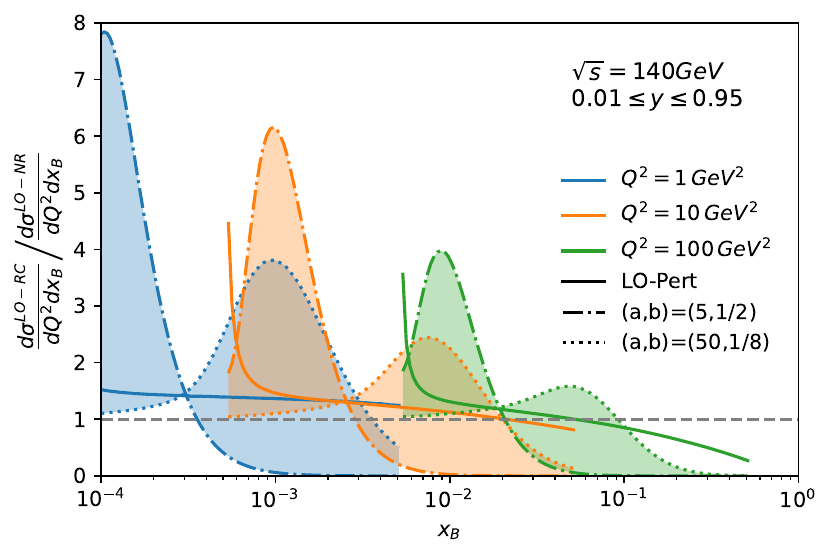} \\
		(a) &{\hskip 0.1in} (b) \\
		\includegraphics[width=0.45\textwidth]{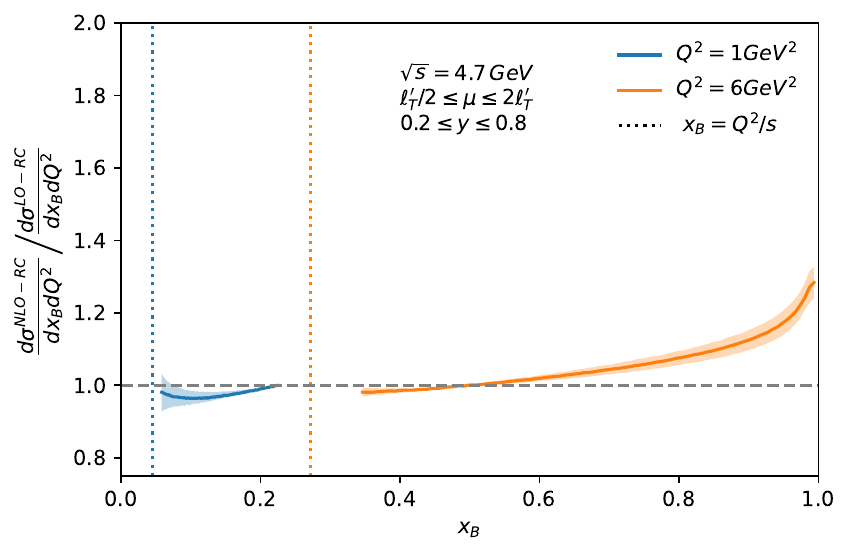} 
	          & {\hskip 0.06in}
		 \includegraphics[width=0.45\textwidth]{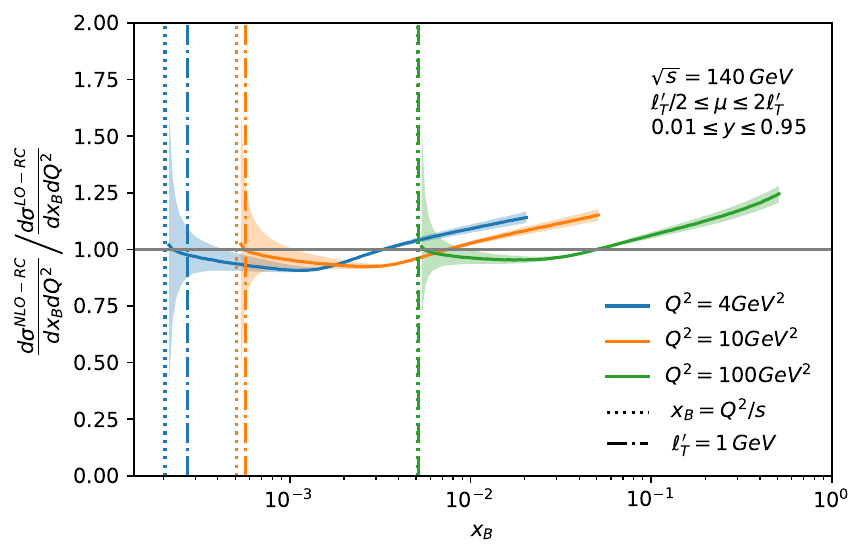} 
                  \\
		 (c) &{\hskip 0.1in} (d)
	\end{tabular}
	\caption{Impact of collision-induced QED radiation to DIS at $\sqrt{s}=4.7$~GeV (a) and $\sqrt{s}=140$~GeV (b);
	              and ratio of NLO over LO QED contribution to DIS at 
	              $\sqrt{s}=4.7$~GeV (c) and $\sqrt{s}=140$~GeV (d). 
	              }
	\label{fig:qed_rad}
\end{figure}

However, both LDFs and LFFs are nonperturbative and should vanish as $\xi$ and $\zeta$ go to 1 since we expect a vanishing probability to find an electron carrying 100\% of its parent electron's momentum once the radiation of a photon is allowed, while these distributions might be peaked at a much larger momentum fraction different from the behavior of light-flavor PDFs. To test the impact of collision-induced QED radiation to DIS cross sections due to the uncertainty of LDFs and LFFs, we parametrize the LDFs and LFFs at an input scale ($\mu_0\sim$~GeV) as 
\begin{equation}
f_{e/e}(\xi,\mu_0^2) = 
\frac{\xi^a (1-\xi)^b}{B(1+a,1+b)}\, ,
\quad\quad\quad
D_{e/e}(\zeta,\mu_0^2) = 
\frac{\zeta^\alpha (1-\zeta)^\beta}{B(1+\alpha,1+\beta)}\, .
\label{eq:ldflff}
\end{equation}
With the Beta function, these parametrized LDFs and LFFs are normalized to have the same integrated probability as those in Eqs.~(\ref{eq:ldf}) and (\ref{eq:lff}).  For the simplicity of the current discussion, we choose  $f_{e/e}(\xi,\mu_0^2) = D_{e/e}(\zeta,\mu_0^2)$ with two sets of parameters: $(a,b)=(50,1/8)$ or (5, 1/2), as shown in the inset of Fig.~\ref{fig:qed_rad}(a) to cover a wide range of LDFs and LFFs.  In Fig.~\ref{fig:qed_rad}(a)-(b), we plot the DIS cross section $d\sigma_{\ell P\to \ell' X}^{\rm LO-RC}$ calculated with these model LDFs and LFFs and $\widehat{H}^{(2,0)}_{ea\to eX}$ at two collision energies over $d\sigma_{\ell P\to \ell' X}^{\rm LO-NR}$.  The solid lines are calculated with the perturbative LDFs and LFFs in Eqs.~(\ref{eq:ldf}) and (\ref{eq:lff}), respectively, while the shaded area represent the range of cross sections given by the two sets of LDFs and LFFs in Eq.~(\ref{eq:ldflff}).  
In Fig.~\ref{fig:qed_rad}(c)-(d), we plot the ratio of the DIS cross sections calculated with NLO over LO hard parts in QED with the set of LDFs and LFFs having (50,1/8) parameters.  The ratio is not very sensitive to the evolution of the distributions~\cite{CQWZ:nlo:qed}.  

%----------------------------------------------------------------
% Figure: Impact of QED radiation
%----------------------------------------------------------------
%----------------------------------------------------------------

%==================================================================
\section{Summary and Outlook}
\label{sec4:summary}

We calculated inclusive lepton-hadron DIS in a joint QCD and QED factorization approach with the hard parts evaluated in both LO and NLO in QED.  We emphasize that like PDFs, LDFs and LFFs are universal and nonperturbative, and their scale dependence depends on evolution kernels evaluated in both QCD and QED.  We found that the impact of QCD on LDFs and LFFs is so much stronger than the impact of QED on PDFs~\cite{QW:evo}.  From Fig.~\ref{fig:qed_rad}, we find that the collision-induced QED radiation has a very strong impact on the size of DIS cross sections which is very sensitive to the shape of LDFs and LFFs.   We also find that impact of NLO QED corrections to the hard parts can generate sizable and nontrivial $x_B$- and $\sqrt{s}$-dependent corrections as shown in Fig.~\ref{fig:qed_rad}(c)-(d). The shaded area in Fig.~\ref{fig:qed_rad}(c)-(d) represents the uncertainty from varying the factorization scale $\mu$ for $\ell'_T/2\leq \mu \leq 2\ell'_T$ with $\ell'_T$ being the transverse momentum of observed lepton.  With the $(\ell'_T)^2=Q^2 (1-y)$, the scattered lepton $\ell'_T$ could be much less than the $Q$ if $y$ is too close to 1 as indicated by the dotted vertical lines in Fig.~\ref{fig:qed_rad}(c)-(d), and $(\ell'_T)^2$ might be a better hard scale than the Lorentz invariant $Q^2$.  

With the universal LDFs and LFFs, the joint QCD and QED factorization approach to the lepton-hadron scattering has no free-parameters other than the factorization scale.  Like QCD factorization, it provides a robust approach to improve the precision of our calculations and predictions, and to explore new observables as well as the dependence on collision kinematics.  Predictions come when we compare different observables depending on the same LDFs, LFFs and PDFs.

%================================================================
% acknowledgement
%-------------------------
\vspace{3mm}
%================================================================
This work is supported in part by the U.S. Department of Energy (DOE) Contract No.~DE-AC05-06OR23177, 
under which Jefferson Science Associates, LLC operates Jefferson Lab. 

%================================================================================
\bibliographystyle{apsrev}
\bibliography{reference}
%================================================================================

%==================================================================
\end{document}